# Title

Signatures of a Spin-Active Interface and Locally Enhanced Zeeman field in a Superconductor-Chiral Material Heterostructure

## Authors

Cliff Chen[1†], Jason Tran[1†], Anthony McFadden[2], Raymond Simmonds[2], Keisuke Saito[3], En-De Chu[1], Daniel Morales[1], Varrick Suezaki[1], Yasen Hou[4,5], Joe Aumentado[2], Patrick A. Lee[4], Jagadeesh S. Moodera[4,5] and Peng Wei[1*]

## Affiliations

[1.] Department of Physics and Astronomy, University of California, Riverside, CA 92521, United States

[2.] National Institute of Standards and Technology, Boulder, CO 80305, United States

[3.] Rigaku Americas, A Division of Rigaku Americas Holding, The Woodlands, TX 77381, United States

[4.] Department of Physics, Massachusetts Institute of Technology, Cambridge, MA 02139, United States

[5.] Francis Bitter Magnet Laboratory, and Plasma Science and Fusion Center, Massachusetts Institute of Technology, Cambridge, MA 02139, United States

* peng.wei@ucr.edu
† These authors contributed equally to this work

## Abstract

A localized Zeeman field, intensified at heterostructure interfaces, could play a crucial role in a broad area including spintronics and unconventional superconductors. Conventionally, the generation of a local Zeeman field is achieved through magnetic exchange coupling with a magnetic material. However, magnetic elements often introduce defects, which could weaken or destroy superconductivity. Alternatively, the coupling between a superconductor with strong spin-orbit coupling and a non-magnetic chiral material could serve as a promising approach to generate a spin active interface. In this study, we leverage an interface superconductor, namely induced superconductivity in noble metal surface states, to probe the spin active interface. Our results unveil an enhanced interface Zeeman field, which selectively closes the surface superconducting gap while preserving the bulk superconducting pairing. The chiral material, i.e. trigonal tellurium, also induces Andreev bound states (ABS) exhibiting spin polarization. The field dependence of ABS manifests a substantially enhanced interface Landé g-factor ($g_{eff}$ ~ 12), thereby corroborating the enhanced interface Zeeman energy.

## Teaser

A spin-active interface between a chiral material and a superconductor with strong spin-orbit coupling has been demonstrated.

**MAIN TEXT**

**Introduction**

Heterostructures combining a structurally chiral material with a superconductor have recently unveiled unique features of induced spin polarization at the superconductor interface(1) and quasiparticle states of magnetic origin(2-5). The underlying mechanism is intricately associated with chirality-induced spin selectivity (CISS), a phenomenon that connects the structural chirality to the orbital angular momentum and spin of electrons. Although the theoretical framework of CISS is evolving, recent models and experimental investigations have underscored the imperative role of spin-orbit coupling (SOC) in facilitating the conversion of chirality-induced orbital polarization of an electron into spin polarization(6-8). In the context of superconductivity, therefore, obtaining an interface superconductor with strong SOC is highly desirable to elucidate the manifestation of CISS, and would be crucial for superconducting spintronics and topological superconductivity(9). Further, because CISS is an interface phenomenon, the quality of the interface is critical. In contrast to previous reports, where the interface is constructed ex-situ by integrating a superconductor with organic chiral molecules(1-5), an interface constructed under ultra-high vacuum (UHV) is desired. Such an interface may substantially enhance the Zeeman energy at the surface of a superconductor, which could serve as a candidate for creating unconventional quasiparticles such as Majorana bound states(10,11). The spin-active interface may also lead to well-defined spin-polarized quasiparticle bound states, i.e. Andreev bound states (ABS), suitable for constructing an Andreev spin qubit.

**Results**

By leveraging proximity effect, we induce superconductivity into a (111)-oriented Au layer, i.e. Au(111), to obtain a surface superconductor with strong SOC. We make Au(111) thin with the aim to acquire a full superconducting gap from an epitaxially grown superconductor layer underneath (Fig 1a). The Au(111) layer is expected to host Shockley surface states, which become superconducting due to proximity effect and may be used to sense any local Zeeman field once an interface is formed on top(12,13). We synthesize high quality Au(111)/Nb bilayers (Methods and SM section-I). The crystallinity, sharp interface, and ultra-thin layers are confirmed by grazing incidence X-ray (GIXRD) and in situ electron diffraction characterizations (Fig 1a and SM section-I). The results demonstrate a substantially improved layer quality compared to the previously reported Au(111)/V materials(12-14). Because of strong SOC, the layers break the Pauli paramagnetic limit. Fig 1b shows the in-plane upper critical field ($H_{C,\parallel}$) of Au(111)/Nb samples with varying Nb and Au thicknesses. When both Nb and Au(111) are thin, we obtain $H_{C,\parallel} > H_P$. Here $H_P$ is the Pauli paramagnetic limit defined as $H_P = \frac{\Delta}{\sqrt{2}\mu_B}$ ($\Delta$ the superconducting gap, $\mu_B$ Bohr magneton) to account for the pair breaking energy when the magnetic field acts on the spin of the quasiparticles and when SOC is negligible(15). When SOC is considered, we adopt the implicit relation describing the pair breaking mechanism(15,16):

$$ln\frac{T_C}{T_{C0}} = \psi\left(\frac{1}{2}\right) - \psi\left(\frac{1}{2} + \frac{\alpha}{2\pi k_B T_C}\right) \quad (1)$$

Here $\alpha$ is the pair-breaker strength (or pair breaking energy) defined as $2\alpha = \frac{\hbar}{\tau_k}$ with $\tau_k$ the lifetime of the Cooper pairs. $T_{C0}$ is $T_C$ under zero applied field and $\psi$ is the digamma function. In the presence of a planar field $H$ and spin-orbit scattering, $2\alpha = \hbar\tau_{SO}(\frac{2\mu_B H}{\hbar})^2$ holds with $\tau_{SO}$ denoting the spin-orbit scattering time. Thus, $\alpha = \frac{2\mu_B^2 H^2}{\Delta_{SO}}$ with $\Delta_{SO} = \frac{\hbar}{\tau_{SO}}$ defining the SOC splitting energy. Using Eq. (1), we fit the data in Fig 1b and obtain $H_{C,\parallel}/H_P \sim 1.4$ for Au(111) (7.5 Å)/Nb (4 nm) with $\Delta_{SO} \sim 1.4$ meV. Because the heterostructure is consist of both Au(111) and Nb and the SOC of Nb is small, the fitted value of $\Delta_{SO}$ may depend on Nb thickness. We expect larger $\Delta_{SO}$ when Nb is thinner. We would also like to note that breaking the Pauli paramagnetic limit would otherwise not be possible if the thin (4 nm) Nb layer were not protected by the air inert Au(111) layer from oxidization.

To create an interface between a chiral material and Au(111), we explore low gap semiconductors, such as selenium (Se) or tellurium (Te), as a tunnel barrier. These materials, i.e. Se and Te, are known to have a trigonal phase, which is structurally chiral(17). We carry out modified point contact measurements using normal metal leads and barriers with a variable thickness to probe the induced superconductivity in Au(111) (Methods and Fig 1c). The $dI/dV$ spectra are described by the Blonder-Tinkham-Klapwijk (BTK) formalism (SM section-II)(15,16). We show that the spectra are modifiable upon changing the thickness of the tunnel barrier – evolving towards the tunneling dominated regime when the thickness increases, for example in Se barrier in Fig 1d (right). The BTK fitting results are given in Table S1 and discussed in SM section-II. However, because of the amorphous growth of the Se barrier, the quasiparticle lifetime broadening ($\Gamma$) is large (SM section-II), which obscures any subtle tunneling features that are narrow in energy.

Using an epitaxially grown Te barrier (Fig 1d left), we obtain tunneling spectra with a dramatically reduced $\Gamma$, which is limited by temperature $k_B T$ (SM section-II). Because $\Gamma$ is reduced, sharp tunneling features reminiscent of two superconducting gaps are observed in a thick Au(111) layer (15 nm) (Fig 1d left). The energy of the larger gap agrees with the bulk superconducting gap estimated from $\Delta \sim 1.76 k_B T_C$ according to BCS model(15) and the measured $T_C = 6.3$ K. We fit Fig 1d (left) using a sum of two BTK $dI/dV$ conductance, namely a two-gap approach (SM section-III), to include both the bulk ($\Delta_B$) and surface ($\Delta_S$) gaps. The fit reproduces the data over a wide range of $V_{bias}$, yielding $\Delta_B \sim 0.85$ meV and $\Delta_S \sim 0.33$ meV, which exactly match the locations of the $dI/dV$ peaks in Fig 1d (left). Although signatures of $\Delta_S$ has been observed in Au(111) before(13), here $\Delta_S$ stands out much more pronounced. In fact, Fig 1d (left) precisely reproduces Potter & Lee's theory of induced surface superconductivity in Au(111)(18). Compared to the previous work where $\Delta_S \sim 0.38$ meV(13), here $\Delta_S \sim 0.33$ meV is comparable.

When a planar field $\mu_0 H$ is applied, i.e. a magnetic field applied parallel to the thin film surface, $\Delta_B$ and $\Delta_S$ evolve with $\mu_0 H$ (Fig 2a). The gaps get reduced and are accompanied by a weakening of the coherence peaks. Using the two-gap BTK model fitting, we reproduce the experimental data under all $\mu_0 H$ (Fig 2a). The fitting in Fig 2a further confirms the two-gap picture, i.e. the coexistence of $\Delta_B$ and $\Delta_S$. To better demonstrate how $\Delta_B$ and $\Delta_S$ evolve with $\mu_0 H$, we plot the $dI/dV$ spectra as a function of $V_{bias}$ and $\mu_0 H$ in a density plot (Fig 2b). The closing of $\Delta_S$ is directly tied to the fill-up of the superconducting gap in the low $V_{bias}$ regime (Fig 2b). The coherence peaks corresponding to $\Delta_B$ also move in towards lower energies as $\mu_0 H$ increases. To quantify such a field dependence, we plot $\Delta_S$ and $\Delta_B$ against $\mu_0 H$ in Fig 2c. Here, the values of $\Delta_S$ and $\Delta_B$ are deduced from the two-gap BTK fitting according to the data in Fig 2a. There is a sharp contrast on their field dependence. $\Delta_S$ vs. $\mu_0 H$ follows a square root dependence and is well fitted according to $\Delta_S \sim \sqrt{1 - (\frac{H}{H_C})^2}$ (Fig 2c), which describes the gap closing of a superconductor in the thin film limit, i.e. thickness $\ll$ penetration depth(15). This agrees with the two-dimensional nature of $\Delta_S$. The square root fitting also outlines the gap in the low $V_{bias}$ regime in Fig 2b, which confirms that the fill-up of the gap corresponds to the closing of $\Delta_S$. On the other hand, $\Delta_B$ vs. $\mu_0 H$ does not follow the square root dependence (Fig 2c) – consistent to the bulk nature of $\Delta_B$. From the fitting, we extract the surface gap ($\Delta_S$) critical field ~ 2.1 T, whereas the critical field for $\Delta_B$ is larger (Fig 2c). Interestingly, although $\Delta_S < \Delta_B$ and one may expect smaller critical field for $\Delta_S$, the critical field (~ 2.1 T) is too small compared to the Pauli paramagnetic limit using $H_P = \frac{\Delta_S}{\sqrt{2}\mu_B}$ and $\Delta_S \sim 0.33$ meV (Fig 1d left and Fig 2a). This may suggest that $\Delta_S$ is experiencing a stronger depairing mechanism. However, to confirm it, one need to make $\Delta_S$ and $\Delta_B$ comparable. Signatures of $\Delta_S$ is also observed in another sample with a 10 nm Au(111) (SM section -V).

To confirm that $\Delta_S$ and $\Delta_B$ are subject to different depairing mechanisms, we obtain $\Delta_S \approx \Delta_B$ in a thinner (5 nm) Au(111) sample. $\Delta_S \approx \Delta_B$ allows us to compare their field dependence (or depairing mechanism) side-by-side. Similarly, we carry out $dI/dV$ measurements across an epitaxially grown Te barrier and control the Te thickness so that tunneling is dominated. At $\mu_0 H = 0$ T (Fig 3a), $dI/dV$ shows one pair of coherence peaks. However, the evolution of the $dI/dV$ spectra under $\mu_0 H$ cannot be explained by a single superconducting gap. At low fields, the gap is very sensitive to $\mu_0 H$. The gap gets quickly filled-up accompanied by a transition of the $dI/dV$ spectra from a U-shape gap (Fig 3a, 3b) to a V-shape gap ($\mu_0 H \sim 3$ T, Fig 3c). At higher fields (after the transition, or $\mu_0 H > 3$ T), the gap widens up again (Fig 3d) and evolves back towards a U-shape gap compared to Fig 3c. Interestingly, at high fields, the gap becomes robust and is nearly unresponsive to $\mu_0 H$, which is demonstrated by an overlapped $dI/dV$ spectra in a wide range of fields from $\mu_0 H \sim 5.5$ T to 8.0 T (Fig 3d). This implies that the low field gap and the high field gap have different origins. Further, the transition between U- to V-shape $dI/dV$ spectra signifies a non-concurrent closing of multiple superconducting gaps, which has also been reported elsewhere, for example in Au(111) with EuS islands(12). Therefore, Fig 3a-d agree with the coexistence of two superconducting gaps in Au(111). Here, we note that the U- to V-shape transition is not due to other sub-gap states, which are resolvable in our high resolution $dI/dV$ measurements, for example the Andreev bound states (ABS) resolved in Fig 3e and Fig 4a. To better demonstrate the non-concurrent gap closing, we carry out detailed $dI/dV$ measurements as a function of $\mu_0 H$ with a fine step field increment (Fig 3e). The density plot in Fig 3e clearly outlines the fill-up and closing of the low field gap, which resembles that in Fig 2b. However, because $\Delta_S \approx \Delta_B$ at $\mu_0 H = 0$ T, the fast closing of $\Delta_S$ suggests that $\Delta_S$ is subject to a stronger depairing mechanism. To further prove that the high field gap (Fig 3d) corresponds to the bulk gap (or $\Delta_B$), we carry out simultaneous transport measurements (Fig 3f). The fact that the sample resistance stays precisely at zero throughout the full scan of $\mu_0 H$ (Fig 3f) confirms that the bulk sample is superconducting.

The stronger depairing of $\Delta_S$ shown in Fig 3e is unusual for a two-dimensional superconductor under a planar $\mu_0 H$. To describe the pair breaking, we recall Eq 1. Because the pair breaking energy $2\alpha = \frac{\hbar}{\tau_k} = \frac{(2\mu_B H)^2}{\Delta_{SO}}$, we rewrite $\alpha = \frac{E_Z^2}{\mu_B H_{SO}}$ with $\Delta_{SO} = 2\mu_B H_{SO}$ and $E_Z = \mu_B H$. Here $E_Z$ is the corresponding Zeeman energy due to the applied field $\mu_0 H$, whereas $H_{SO}$ is introduced as an effective field to account for SOC. Considering that superconductivity is destroyed at $\alpha \sim \Delta$, one would expect $\Delta_S$ and $\Delta_B$ to have similar critical field or close at similar $\mu_0 H$ (or $E_Z$) because $\Delta_S \approx \Delta_B$ at $\mu_0 H = 0$ T (Fig 3). When SOC is taking into account, we expect the surface/interface critical field $H^S_{C,\parallel}$ (corresponds to $\Delta_S$) is even higher, because stronger SOC is expected at the interface according to Fig 1b. To estimate the bulk critical field ($H^B_{C,\parallel}$), we fit the transport data according to Eq 1 (Fig 3f inset), which yields $H^B_{C,\parallel} \sim 8.1$ T at 300 mK. To estimate $H^S_{C,\parallel}$, we note that it shall be reached shortly after the fill-up of the low-field gap across the transition at $\mu_0 H \sim 3$ T (compared to Fig 2b). Therefore, we have $H^S_{C,\parallel} < H^B_{C,\parallel}$. Such a stronger depairing on $\Delta_S$ cannot be solely explained by $\mu_0 H$, and one needs to consider other depairing mechanisms for surface/interface.

To explain $H^S_{C,\parallel} < H^B_{C,\parallel}$, we consider a stronger depairing for $\Delta_S$ at the Te/Au(111) interface. Here, we notice that the surface of Au(111) is also the interface between Au(111) and the tunnel barrier Te (Fig 1c), which can be chiral if it is in the trigonal phase. We rewrite $E_Z = \frac{1}{2} g_{eff} \mu_B H$ with $g_{eff}$ the effective Landé $g$-factor, which absorbs all the contributions that enhance the interface $E_Z$. To find out the relationship $E_Z \sim H$, we focus on the low field regime in Fig 3e by expanding the field dependence of the in-gap states and re-plot it in Fig 4a (inset). Thanks to the high-quality epitaxial Te barrier with small $\Gamma$ (SM section-II), subtle in-gap states are resolved. These states reminiscent themselves as a pair of small $dI/dV$ peaks moving with $H$ in a linear way (Fig 4a), which suggests that they could be ABS. Also, the linear field dependence and the crossing of the peaks (Fig 3e and Fig 4a) indicates that the states are spin polarized and the change of energy under a magnetic field is $\propto E_Z$. To do a quantitative analysis, we fit the ABS peaks and plot their energy vs. $H$ (with error bar) in Fig 4a. According to $E_Z = \frac{1}{2} g_{eff} \mu_B H$, we have $\Delta E_Z = \frac{1}{2} g_{eff} \mu_B (\Delta H)$. The fitting in Fig 4a yields $g_{eff} \sim 12$. Such $g_{eff}$, a six-fold increase compared to $g \sim 2$ in either Au(19) or Nb(20), cannot be explained by any bulk properties of the sample. Since ABS is often an interface effect, the large $g_{eff}$ is likely due to interfacing Au(111) with the Te barrier. In fact, a picture considering $g_{eff}$ ~ 12 at Te-Au(111) interface and $g$ ~ 2 in bulk Au(111)/Nb is consistent with the combined experimental results in Fig 2 and Fig 3. Noting that the surface gap closes at $\alpha = \frac{E_Z^2}{\mu_B H_{SO}^S} \sim \Delta_S$ with $E_Z = \frac{1}{2} g_{eff} \mu_B H_{C,\parallel}^S$, we have $H_{C,\parallel}^S \sim \frac{1}{g_{eff}/2} \sqrt{H_P H_{SO}^S}$. Here $H_{SO}^S$ is $H_{SO}$ at the surface of Au(111) (or at Te/Au(111) interface), and $H_P = \frac{\Delta_S}{\sqrt{2}\mu_B}$ as defined. According to the thickness dependence in Fig 1b, $H_{SO}^S > H_P$ is expected at the interface. Therefore, we estimate $H_{C,\parallel}^S > \frac{1}{g_{eff}/2} H_P$. Taking $H_P$ ~ 11 T for $\Delta_S$ ~ 0.9 meV ($\Delta_S \approx \Delta_B$), we obtain $H_{C,\parallel}^S > 1.8$ T – consistent to the fast closing of $\Delta_S$ as shown in Fig 3e. The substantially enhanced Landé $g$-factor, i.e. $g_{eff}$ ~ 12, agrees with a spin active interface, which explains the observed ABS in Fig 3e and 4a(21,22). Such an interface is expected to host spin triplet superconductivity(21,23) and requires a generalized BTK formalism beyond Eq S1 to model(23,24). In fact, the nearly magnetic field independent $dI/dV$ spectra in Fig 3d (or after the closing of $\Delta_S$) is very unusual (Fig 3d), which cannot be simply explained by a superconducting gap following Eq S1 under $\mu_0 H$. Also, because $H_{C,\parallel}^B$ ~ 8.1 T (Fig 3f), one would expect a substantial gap reduction when $\mu_0 H$ is approaching $H_{C,\parallel}^B$. However, Fig 3d shows that the gap is nearly unchanged up to $\mu_0 H$ ~ 8.0 T and even slightly increased at $\mu_0 H$ ~ 8.0 T compared to $\mu_0 H$ ~ 5.5 T. Such an unusual superconducting gap (Fig 3d), robust against $\mu_0 H$, could hint the spin triplet superconductivity.

While we extract $g_{eff}$ using ABS, it is not yet clear why there are spin polarized ABS at $\mu_0 H = 0$ T. Particularly, we notice that no magnetic material with a spontaneous magnetization is present. To answer this question, we study the crystal structure of the epitaxially grown Te layer that is used as a tunnel barrier. Using Raman spectroscopy, we observe distinct Raman peaks at A$_1$ = 121.15 cm$^{-1}$, E$_1$ = 92.21 cm$^{-1}$ and E$_2$ = 140.64 cm$^{-1}$ (Fig 4b), which are characteristic to trigonal Te(25). The vibration modes in correspondence to each of the Raman peaks are illustrated in Fig 4b (inset). The A$_1$ mode (the major peak in Fig 4b) corresponds to the breathing mode of the Te helical chain.

We further carry out in situ angle-dependent reflection high energy electron diffraction (RHEED) measurements on the Te layer (SM section-IV). The results suggest that the six-fold rotation symmetry of Au(111) facilitates the growth of trigonal Te, which also has six-fold rotation symmetry (Fig 4b inset and SM section-IV). Combining Raman and RHEED results, we conclude that the Te barrier layer, when carefully grown, can be an epitaxial trigonal Te. The helical Te chains in a unit cell (Fig 4b inset) may give rise to domains with a distinct chirality. The size of the domain and its crystallinity may depend on the quality of Te and could get enhanced by improving the epitaxial growth. Because the junction has a microscopic area (Methods), it may help probing a local region of trigonal Te with a distinct chirality on average. Also, the ABS is a subtle interface phenomenon, therefore it may sensitively depend on the interface quality and the chiral domains of Te. Nevertheless, the combination of chiral Te and a superconductor with strong SOC (Fig 1b and 1c) satisfies the needed components for CISS(6-8). Compared to other work where $g$ ~ 2 is reported(2,3), we demonstrate a much larger $g_{eff}$ ~ 12, which highlights the spin active interface and directly points towards CISS. Here, we note that the high-quality interface we created in situ under UHV could be a key factor leading to the substantially enhanced CISS phenomena, whereas in other work the interface was often created in an ex-situ way by the adsorption of organic chiral molecules on the surface of a superconductor(2-5).

To further demonstrate the superconductor heterostructure Au(111)/Nb has excellent microwave (RF) performance and can potentially be used as a component for superconducting qubits, we fabricate resonators using ultra-thin Au(111)/Nb layers (Fig 4c). We follow the same coplanar waveguide (CPW) design (Methods) as those used in conventional Nb resonators, which are often consisted of Nb layers that are one order of magnitude thicker. We carry out superconducting resonator measurements at 35 mK and find out that Au(111) limits the defect formation, i.e. the formation of uncontrolled Nb oxides, in ultra-thin Nb. It has been shown that the native oxides of Nb are the main sources responsible for two-level defects (TLS), i.e. possible trapping sites for quasiparticles, and additional dielectric losses in commonly used transmon qubits, which causes qubit relaxation or decoherence(26,27). Using resonators fabricated from Au(111)/Nb (Methods), Figure 4d shows an example of the sharp microwave resonance peaks observed in transmission near a RF frequency ~ 5.747 GHz at 35 mK temperature in one of the resonators. The resonance peak is well-defined with a Lorentzian line shape, which is fit using the established diameter correction method (DCM)(28) to extract internal ($Q_i$) and external ($Q_e$) quality factors. An internal quality ($Q$)-factor $Q_i > 10^6$ is achieved when the averaged microwave photon number $\langle n \rangle$ is large (Fig 4e). When $\langle n \rangle$ is reduced towards the single photon limit (Fig 4e), $Q_i$ decreases to values between 2 - $4 \times 10^5$ due to the contribution of saturable TLS to RF power loss(29). We note that the narrow conductor/gap dimensions in our CPW resonator design will result in lower $Q_i$ values than designs with larger dimensions which dilute the participation of surfaces and interfaces (Methods). The $Q_i$ values observed near and below single photon powers in these chips are comparable to values obtained in conventional resonators with one order of magnitude thicker Nb films (100 nm) based on similar CPW design on sapphire (SM section-VI and Fig S7). This suggests that these high-quality Au(111)/Nb structures perform well even for these ultra-thin (12nm) Nb layers and are suited for use in superconducting qubits such as transmons. In addition, the high $Q_i$ values observed suggests Au(111) limits the defects in Nb and thereby resulting minimal in-gap states.

**Discussion**

In conclusion, we have demonstrated an enhanced Zeeman field at Te(trigonal)/Au(111) interface giving rise to $g_{eff}^S$ ~ 12. Such an enhanced interface Zeeman field closes the surface superconducting gap of Au(111) while maintaining the bulk superconductivity, thereby providing an alternative route to creating a unique surface/interface superconductor. One may exploit such locally enhanced Zeeman field to construct a nanostructured topological superconductor (TSC) by patterning, for example, nanowires of trigonal Te over Au(111)(10,11). The Te layer may at the same time serve as a high-quality tunnel barrier – making it advantageous in fabricating multi-terminal planar tunneling devices involving TSC(30,31). An epitaxially grown Te layer in its trigonal form on Au(111)/Nb could also be useful for superconducting spintronics. Also, we have demonstrated a substantially enlarged $\Delta_S$ (Fig 3), several fold of increase compared to the prior report(13), achieving $\Delta_S \approx \Delta_B$ in 5 nm Au(111) grown on Nb. Theory simulations have shown that a larger surface superconducting gap may result in a larger topological gap giving rise to a robust TSC(10,12). Moreover, $\Delta_S \approx \Delta_B$ ensures that the whole Au(111) layer is fully gapped, which limits microwave loss due to quasiparticle excitations across the gap, as well as reducing TLS defects that lead to dielectric losses. These are essential for applying Au(111)/Nb in microwave resonators and superconducting qubits.

**Materials and Methods**

The Au(111)/Nb bilayers are grown using molecular bean epitaxy (MBE). The Nb layer thickness can be controlled down to a few nanometers and the Au(111) layer thickness down to a few angstroms. Following the growth, Au(111)/Nb samples are routinely taken out of the UHV chamber for transport and tunneling measurements in an Oxford 3He Heliox® cryostat. The point contact spectroscopy is carried out using a normal metal (gold) wire touching the sample surface forming a junction with a controlled tunnel barrier layer (Te or Se) added. The point contact wire is made in a way to allow four-terminal differential conductance ($dI/dV$) measurement through the junction (Fig 1c inset). The junction usually has an area of ~ 20 μm in diameter. $dI/dV$ measurement is performed at a temperature ~ 300 mK with a modulation voltage $\leq k_B T$.

The Au(111)/Nb CPW resonators are fabricated following a standard design consisting of eight inductively coupled quarter wave resonators in a "hanger" configuration with resonances falling between 4 - 8 GHz using standard photolithography steps and Ar ion milling. The CPW design chosen has conductor/gap dimensions of 6μm /3μm. These narrow dimensions are chosen to increase the participation of the surfaces and interfaces thus making the loss measurement more sensitive to their contribution(32). After resonator fabrication, the chip was installed in a Au-plated

Cu sample package using Al wire bonds. RF characterization of frequency multiplexed CPW resonators was performed in a dilution refrigerator having a base temperature of 35 mK. Each chip having dimensions 7.5 mm × 7.5 mm contains 8 frequency-multiplexed ¼ wave hanger resonators with a central feedline. Power dependent transmission ($S_{21}$) measurements were performed at base temperature spanning powers corresponding to ~$10^6$ to ~0.1 photon occupation in the resonator. Each resonator is measured using a vector network analyzer, and the Lorentzian spectra are fit using the established diameter correction method (DCM) to extract internal and coupling quality factors ($Q_i$ and $Q_e$) as a function of photon number(28). The signal and amplifier chain used is similar to that outlined previously(32) with amplifiers installed at 4 K and at room temperature. A Josephson parametric amplifier installed at the mixing chamber (35 mK) was used for all low power measurements (below ~ $10^3$ photon occupation).

For sample structure characterizations, ex-situ X-ray based analyses are performed using Rigaku GIXRD system. In situ reflection high energy electron diffraction (RHEED) is carried out using a 7.5 mW electron beam. Ex-situ Raman spectroscopy is carried out using a Horiba LabRam system, in which a 532 nm unpolarized excitation laser with a 100μm spot size was used to scan the films.

## Acknowledgments

We thank Jing Shi for the support on 3He millikelvin temperature measurements. We thank Corey Rae McRae for the discussions on superconductor resonators. CC, JT, EC, DM, VS and PW would like to acknowledge the support from NSF CAREER DMR-2046648. PAL, JSM and PW would like to acknowledge the support from NSF Convergence Accelerator Track-C ITE-2040620. JSM and PW would like to acknowledge the Lincoln Lab Line fund. YH and JSM would like to acknowledge National Science Foundation NSF-DMR 1700137 and 2218550, Office of Naval Research N00014-20-1-2306, Army Research Office W911NF-20-2-0061 and DURIP W911NF-20-1-0074, and the Center for Integrated Quantum Materials NSF-DMR 1231319. PAL would like to acknowledge the support by DOE office of Basic Sciences Grant No.DE-FG0203ER46076.

## Author contributions:

CC, JT and VS synthesized the bilayer thin films. CC and JT carried out tunneling and transport measurements. KS performed GIXRD measurements and data analysis. DM and VS fabricated the resonator devices. AM and RS performed microwave absorption measurements. EC performed Raman measurements and analysis. PW oversaw the project and designed the research with the discussions from JA, PL, and JSM. All authors contributed to writing the manuscript.

## Competing interests:

Peng Wei is the inventor on a technical disclosure to be submitted by the University of California, Riverside. All authors declare no other competing interests.

## Data and materials availability:

All data needed to evaluate the conclusions of the paper are present in the paper and/or the Supplementary Materials. The data will be publicly accessed on Dryad under the link:

https://doi.org/10.5061/dryad.4xgxd25jf
Page 15 of 27

**Figures and Tables**

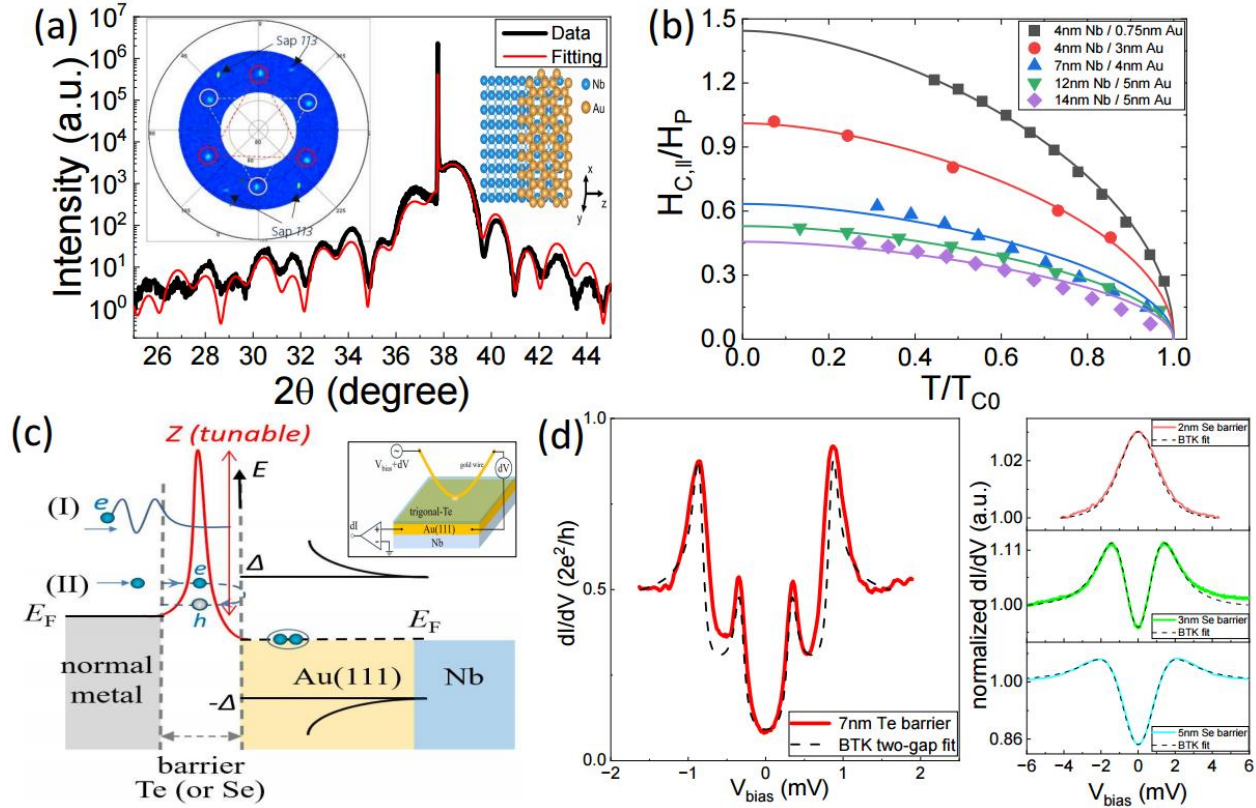

**Figure 1 | Superconductor with strong spin-orbit coupling (SOC), signatures of the induced surface superconducting gap, and tunneling to Au(111) surface using tunnel barriers with adjustable properties** (a) GIXRD $2\theta - \omega$ scan. The main diffraction peak is at around $2\theta \sim 38.5°$, which manifest itself as an overlapped Nb(110) and Au(111) peaks. The fringes/beatings reflect the thicknesses of the layers and are fitted using dynamical simulation fittings (red curve). Inset (left) shows the X-ray pole figure scan confirming (111)-oriented gold surface. The schematic layout of the Au(111) and Nb(110) is shown in the inset (right). (b) The in-plane upper critical field $H_{C,\parallel}$ is plotted against temperature in samples with a variety of Au and Nb thicknesses. $H_{C,\parallel}$ is normalized to $H_P$ and $T$ is normalized to the critical temperature at zero field $T_{C0}$. Reducing the thicknesses of both Au and Nb increases $H_{C,\parallel}/H_P$. (c) A schematic of the quasiparticle reflection processes across the interface including ordinary reflection (tunneling) and Andreev reflection (SM section-II). A structure layout of the four-terminal point contact device is shown in the inset (Methods). (d) Left: The $dI/dV$ spectrum measured using tellurium (Te) (7 nm) as a barrier on a Au(111)( 15 nm)/Nb sample. The data is fitted (dotted line) using a two-gap BTK approach (SM section-II and -III). Right: Adjusting the barrier height on Au(111)/Nb using a similar low-gap semiconductor, i.e. selenium (Se), with a varying Se thickness: 2nm (top), 3nm (middle) and 5nm (bottom). The dotted line is the fit to the BTK theory with fitting results given in Table S1.

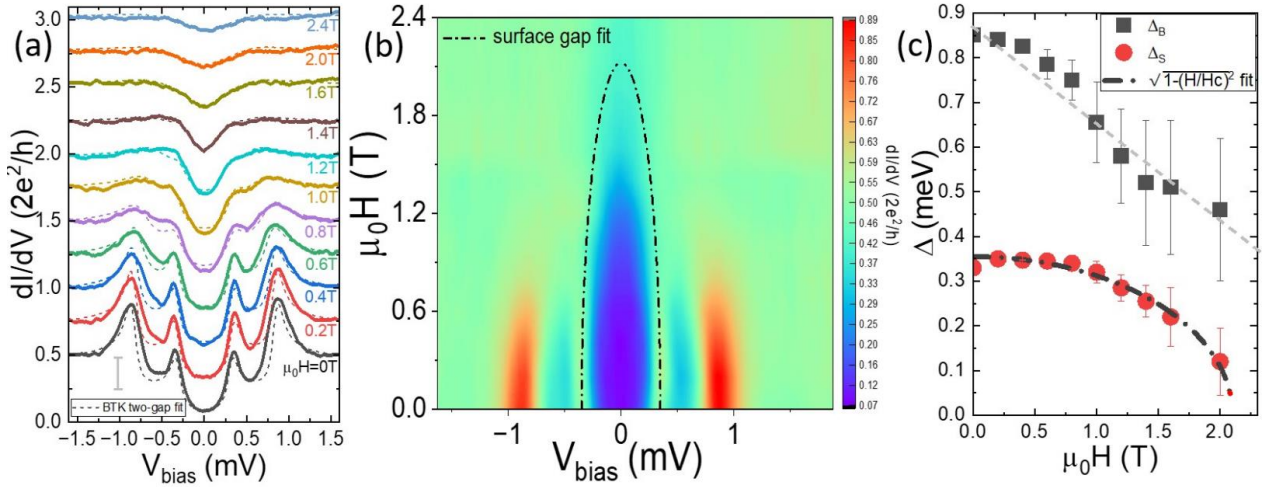

**Figure 2 | Evolution of the bulk $\Delta_B$ and surface $\Delta_S$ superconducting gap under a planar magnetic field.** (a) The fit to the $dI/dV$ spectra using the two-gap BTK approach for various magnetic fields applied parallel to the Au(111)/Nb thin film sample. The sample (also shown in Fig 1d, left) has a 15 nm thick Au(111) layer and 7 nm Te barrier. The gaps $\Delta_S$ and $\Delta_B$ are deduced from the fitting. The spectra are vertically shifted for clarity with a scale bar indicating 0.5 $e^2$/h. (b) The density plot of the $dI/dV$ data as a function of $V_{bias}$ and $\mu_0 H$. The surface gap closes soon after a substantial fill-up of the superconducting gap (region in blue color). The dash-dot line shows the dependence of $\Delta_S$ vs. $\mu_0 H$ following the fitting results in (c). (c) The evolution of $\Delta_S$ and $\Delta_B$ as a function of $\mu_0 H$. Here $\Delta_S$ and $\Delta_B$ are deduced from the fitting in (a). The increased error bar is due to the increased broadening under large fields. The dash-dot line (black) is the fitting results according to $\Delta_S \sim \sqrt{1-(\frac{H}{H_C})^2}$ with the consideration of the error bars. The fitting is consistent with the picture that $\Delta_S$ is the surface gap, which is two-dimensional in nature. The bulk gap $\Delta_B$ does not follow such a field dependence, and the dashed line (grey) is guided by eye. The critical field of $\Delta_S$ is smaller than that of $\Delta_B$.

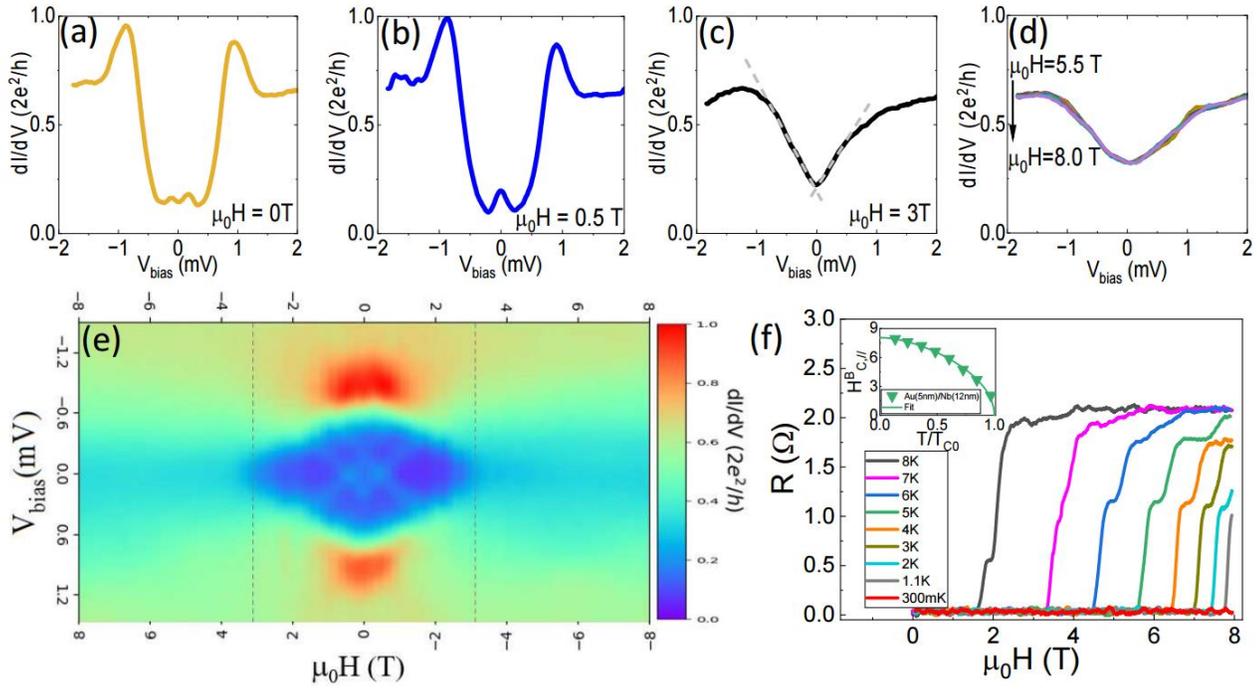

**Figure 3 | Fast closing of $\Delta_S$ and signatures of an enhanced interface Zeeman field. (a) - (c)** Characteristic $dI/dV$ spectra under the planar $\mu_0H$: 0 T, 0.5 T, and 3.0 T in a Au(111)(5nm)/Nb with an epitaxial Te barrier. At 0 T (a), despite the two subtle tunneling peaks near $V_{bias}$ ~ 0 mV (ABS in Fig 4a), a single superconducting gap is observed. As $\mu_0H$ increases, the gap gets narrowed and filled-up accompanied by an evolution from "U-shape" in (a) - (b) to "V-shape" in (c). The dotted line in (c) is guided by eye. **(d)** At high fields, the gap becomes nearly unresponsive to $\mu_0H$, and the $dI/dV$ spectra overlap in a wide range of fields (5.5 T to 8.0 T). Compared to (c), the gap in (d) is widened and evolving back towards the "U-shape". **(e)** $dI/dV$ vs $V_{bias}$ and $\mu_0H$. Two gapped regimes are visible: one (dark blue regime) closes approximately at $\mu_0H$ ~ 3 T, and the other one (light blue regime) survives up to $\mu_0H$ ~ 8 T and beyond (also see d). The transition, i.e. the closing of the dark blue regime, is accompanied by the fill-up of the low field gap (also see a-c). At $\mu_0H$ ~ 0 T, the two gaps overlap. **(f)** The simultaneous measurements of $R$ vs $\mu_0H$ on the sample in (e). The sample is superconducting (red curve) throughout the whole magnetic field range. Inset: bulk critical field $H^B_{C,\parallel}$ vs $T/T_{C0}$. The fitting is based on Eq 1.

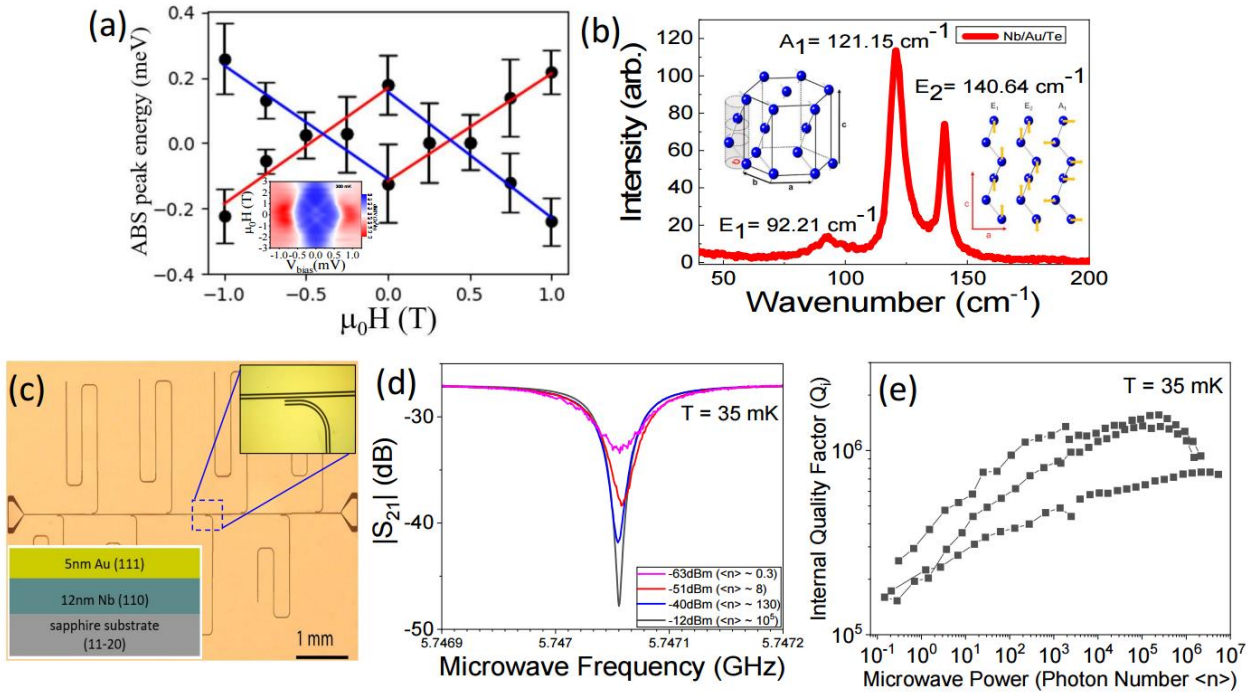

**Figure 4 | ABS demonstrating an enhanced interface Landé *g*-factor and high Q-factor resonators made of ultra-thin Au(111)/Nb.** (**a**) ABS energy vs $\mu_0 H$. The energies, or the peak locations of ABS (see also Fig 3a, 3e), are fitted using Gaussians with uncertainties indicated by the error bar. The peaks shift linearly with $\mu_0 H$ and crosses each other at $\mu_0 H \sim \pm 0.5$ T suggesting that the ABS could be spin polarized. The straight lines are the linear fit with slopes corresponding to the Landé *g*-factor $g_{eff} \sim 12$, which is six times larger than $g \sim 2$. At the crossing point ($\mu_0 H \sim \pm 0.5$ T), a zero bias conductance peak (ZBCP) is formed (Fig 3b). However, the ZBCP is unstable and splits as soon as $\mu_0 H$ increases. The inset shows the low field $dI/dV$ data corresponding to Fig 3e. (**b**) Raman spectroscopy characteristics of the epitaxial trigonal Te tunnel barrier. The trigonal Te peaks ($A_1$, $E_1$ and $E_2$) are observed with no other vibration modes present. The inset shows the crystal structure of trigonal Te and the vibration modes. (**c**) The array of superconductor resonators fabricated using ultra-thin Au(111) (5 nm)/Nb (12 nm) layers (Methods). The inset shows the stack layout of the layers and a zoomed-in image of one resonator. (**d**) Power dependent transmission ($S_{21}$) measurements signifying the resonance peaks ($T = 35$ mK). (**e**) $Q_i$ vs $\langle n \rangle$ for three typical quarter wave resonators of (c).

# Supplementary Materials

## I. Structural and surface/interface characterizations of Au(111)/Nb

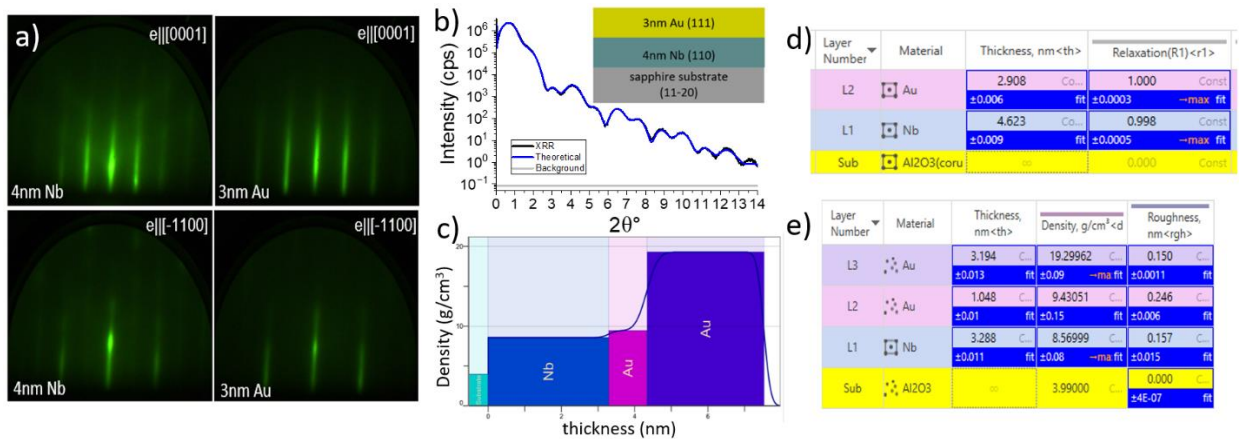

**Figure S1 | Synthesis and structural characterizations of high-quality ultra-thin Au/Nb bilayers:** (a) The RHEED images for both Au and Nb along two characteristic crystalline directions. The sharp streaks confirm high crystal quality and smooth surface/interface. The rotation periodicity demonstrates the crystal symmetry. (b) The XRR results as measured by GIXRD. The black curve is the measured data, and the blue curve is the model fitting. Inset: The stack layout of the Au/Nb heterostructure. (c) The density profile of the heterostructure is plotted against the layer depth showing interface/surface quality. The layer depth profile is extracted by fitting the XRR data in (b). (d) The result of the dynamical simulation fitting to Fig 1a in the main text. The fitted layer thickness matches well with the growth thickness. (e) The numerical fitting results for the XRR data. The layer thickness, density, and roughness are given with errors.

High quality epitaxial Au(111)/Nb thin film bilayers are grown using ultra-high vacuum (UHV) molecular beam epitaxy (MBE). The growth is controlled so that the bilayers have sharp surface/interfaces, which allow ultra-thin Au(111) and Nb layers to be made. In situ reflection high-energy electron diffraction (RHEED) is used to probe the crystallinity and rotation symmetry of the terminated thin film surface during the growth of the multi-layer. Fig S1 (left column) shows the RHEED of the surface of 4nm Nb before the growth of Au(111), and Fig S1 (right column) shows the surface of the 3nm Au(111) after the growth. The sharp diffraction streaks indicate the film is highly epitaxial and the growth is two-dimensional[33]. Upon rotating the azimuthal angle, the RHEED streaks alternate in between two distinct patterns corresponding to the electron beam applied parallel to [0001] and [-1100] crystalline directions (Fig S1 top and bottom rows). This confirms that the Au thin film is (111) textured. The crystal quality of the layer is also confirmed by GIXRD $2\theta - \omega$ scan (main text: Fig 1a) with the Nb(110) and Au(111) diffraction peaks overlapped at $2\theta \sim 38.5°$. The thickness fringes and beatings in Fig 1a are well reproduced using dynamical simulation fittings and the fitting results are shown in Fig S1d. The pronounced fringes in the GIXRD $2\theta - \omega$ scan is a direction consequence of sharp surface/interface in the heterostructure. The fitting yields layer thicknesses that match well with the actual growth thickness.

The surface/interface quality is confirmed by low-angle XRR measurements (Fig S1b). The oscillation and beating in XRR are also a consequence of the sharp surface/interface. Fig S1b shows that the XRR oscillations survive up to a high angle $2\theta \sim 14°$ without being overwhelmed by noise and may continue beyond. Since XRR oscillations are the results of the interference after the X-ray is reflected at the surface and interface of the layers, the higher the angle the oscillations would survive the better quality the surface/interface is[14,34]. Further model fittings (Fig S1c and S1e) have been performed to the XRR data in Fig S1b yielding a surface roughness of Au(111) ~ 1.5 Å – a six-fold improvement compared to the previous Au(111)/V samples[14]. The density profile of the heterostructure is plotted in Fig S1c based on the XRR fitting.

In addition to RHEED (Fig S1a), the (111)-oriented gold surface is confirmed by the GIXRD pole figure scan (Fig 1a inset and Fig S2). The Au(200) peak shows sharp diffraction spots with six-fold rotation symmetry (Fig S2a), which confirms Au(111). Interestingly, the Nb(200) peak demonstrates a two-fold rotation symmetry with a lattice rotated with respect to that of Au(111). From the pole figure scan, one may confirm that both the Au(111) and Nb(110) layers are epitaxially grown on sapphire. The relative rotation of their crystal lattices facilitates epitaxial growth. In the GIXRD reciprocal space mapping (RSM) measurement, the sapphire(300) peak and Au(113) peak are shown in Fig S2d. Based on that, the lattice constants of Au are estimated to be $a = b = c = 4.079$ Å, suggesting that the Au(111) layer grown on Nb(110) is relaxed with negligible strain.

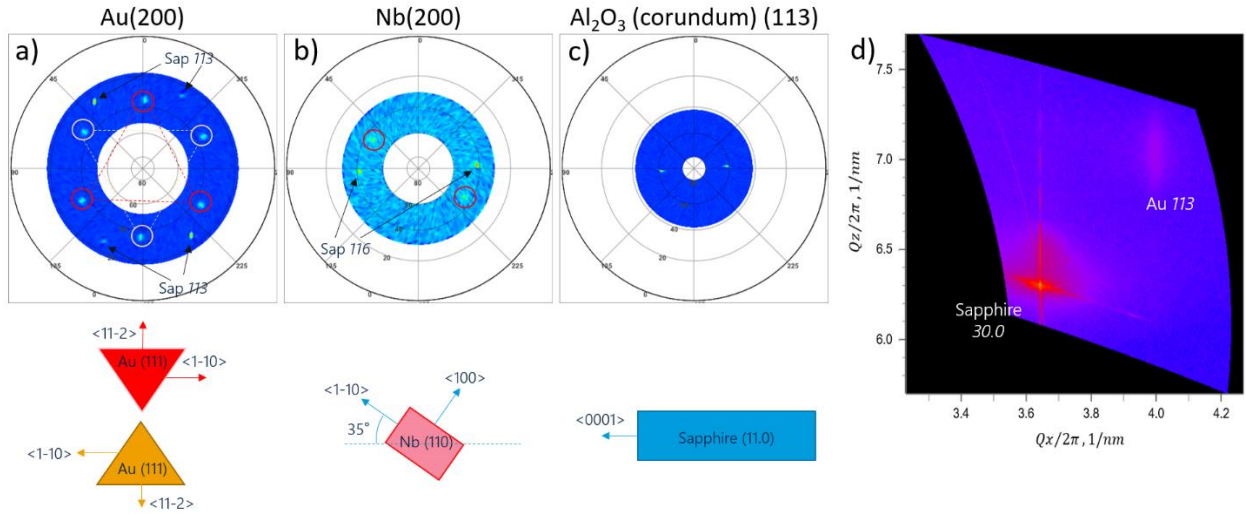

**Figure S2 | Epitaxial relationship between Au(111) and Nb(110): (a)** - **(c)** The pole figure results of Au(200) (a), Nb(200) (b) and Al$_2$O$_3$ (113) (c) measured by GIXRD. The relative position of the diffraction spots illustrates the epitaxial relationship between the layers. The domain with the corresponding symmetry is illustrated below. **(i)** The reciprocal space mapping (RSM) of the layer showing the Au layer is completely relaxed and the amount of strain is small.

## II. Fitting with modified Blonder-Tinkham-Klapwijk (BTK) formalism

The evolution of the $dI/dV$ spectra in Fig 1d (right) reflects the weight change between Andreev reflection and quasiparticle tunneling processes (Fig S3a), which shall follow the BTK formalism[35]:

$$I \propto N(0) \int_{-\infty}^{+\infty} [f(E - eV) - f(E)][1 + A(E) - B(E)]dE \qquad (S1)$$

Here $N(0)$ is the density of states at the Fermi level of the normal metal lead and $f(E)$ is the Fermi-Dirac distribution. The probabilities $A(E)$ describes the Andreev reflection process (enhanced current) and $B(E)$ describes the ordinary reflection of a quasiparticle in a tunneling process (reduced current). While the dependence of $A(E)$ and $B(E)$ on $\Delta$ and barrier strength $Z$ are given by the BTK model[35], we introduce the Dynes $\Gamma$ parameter by substituting $E \to E + i\Gamma$ to account for quasiparticle lifetime broadening[36].

**Table S1 | BTK fitting results on $\Gamma$ and $Z$ according to Eq S1 (data from Fig 1d right)**

| Barrier Thickness | Broadening $\Gamma$ (µeV) | Barrier strength $Z$ |
|---|---|---|
| 2nm (Se) | 590 | 0.3 |
| 3nm (Se) | 330 | 0.8 |
| 5nm (Se) | 750 | 1.5 |

When the thickness of the Se barrier increases, the fitting shows an increased barrier height ($Z$) indicating an enhanced barrier strength (Table S1). However, the quasiparticle lifetime broadening $\Gamma$ stays large, or $\Gamma > k_B T$ (Table S1), implying predominant defect scatterings and short quasiparticle lifetime[36], which agrees with the amorphous growth of Se.

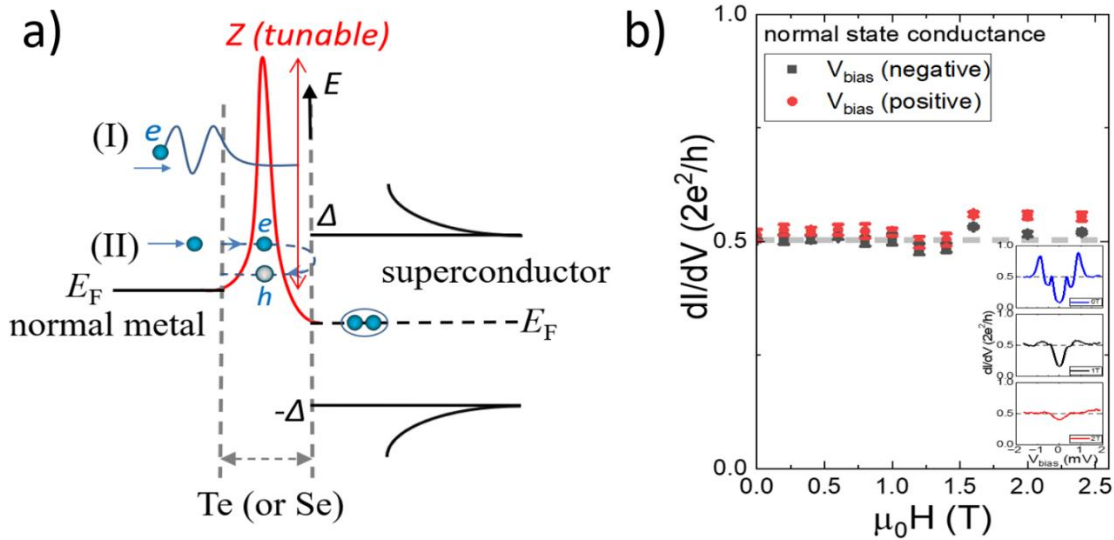

**Figure S3 | Fit the $dI/dV$ conductance with the BTK model and the nearly quantized normal state conductance:** (a) A schematic of the transmission processes in the BTK model. For the point contact measurement. Both the Andreev reflection and quasiparticle reflection (or tunneling) processes play a role in determining the total current (Eq. S1). When the barrier strength $Z$ is large, the quasiparticle tunneling process dominates (Table S1). (b) The normal state $dI/dV$ of Fig 2a plotted against $\mu_0 H$. The conductance stays around $e^2/h$ and is nearly a constant over a wide range of $\mu_0 H$ (from zero field up to 2.5T). This field independent, nearly quantized normal state conductance may indicate a high transparency of the epitaxial Te tunnel barrier.

To enhance the quality of the tunnel barrier or reduce $\Gamma$, we replace Se with Te. We find that Te grows epitaxially on Au(111) down to 1nm (SM section-IV). The $dI/dV$ measurements in Fig 2 are performed on a 15 nm Au(111) layer using a 7nm Te barrier. The thickness of the Te barrier is controlled so that the tunneling process dominates. Using a modified two-gap BTK model fitting (SM section-III), we obtain an order of magnitude reduced $\Gamma$ – resulting $\Gamma \sim k_B T$, i.e. the thermal limit, at a measurement temperature $T \sim 300$ mK. The fitting is shown as a dashed line in Fig 1d left. It also yields $Z \sim 2.6$, which confirms the tunneling dominated regime. Because of the substantially reduced $Z$, the energy resolution of the $dI/dV$ spectra is enhanced, and fine tunneling features are resolved (Fig 1d left). Interestingly, the normal state $dI/dV$ stays around the quantized conductance $e^2/h$ and is nearly constant regardless of $\mu_0 H$ for a wide range of magnetic field from 0T up to 2.5T (Fig S3b). This may suggest a high tunneling transparency.

### III. Two-gap BTK model fitting

To account for a BTK process involving a normal metal lead and a superconductor with two gaps, we take a two-gap approach by modifying Eq S1. Similar two-gap approach has been used previously to address the quasiparticle tunneling process[13]. Here, we expand it to the BTK process by considering that the total current $I$ (Fig 2a) is written as $I = I_B + I_S$. For $I_B$ (or $I_S$), it describes a BTK process involving a normal metal and a superconductor with a gap $\Delta_B$ (or $\Delta_S$), each of which is described by Eq S1. The two-gap model fits the data in Fig 1d (left) and Fig 2a under all the values of $\mu_0 H$. At low bias voltage (or near zero $V_{bias}$), the non-zero $dI/dV$ conductance is due to a finite Andreev reflection process. The $dI/dV$ in the low $V_{bias}$ regime can be well reproduced by the fitting (Fig 1d left), therefore it does not indicate a soft superconducting gap nor a finite density of states withing the gap.

### IV. Growth and structural characterizations of the Te barrier

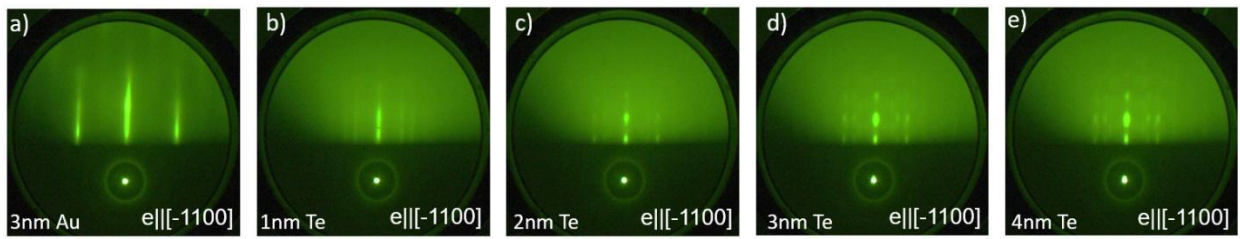

**Figure S4 | RHEED images of the Te tunnel barrier layer during the growth:** **(a)** pristine Au(111) surface before the growth of Te, **(b)** 1nm Te shows clear RHEED streaks, **(c)-(e)** 2nm Te layer starts to show RHEED dots, which becomes more prominent in 3nm and 4nm Te.

Te tunnel barrier is grown immediately following the growth of Au(111) for those samples to be measured by point contact spectroscopy. RHEED is taken in situ during the growth. Fig S4a shows the RHEED streaks of Au(111) before the growth. The RHEED of Te shows streaky diffraction lines when the thickness of Te is 1nm, indicating layer-by-layer Te growth at the thin layer limit. When the Te thickness increases above 2nm, the streak lines covert into dots indicating the formation of islands[33]. This growth mode is consistent with the Stranski–Krastanov (SK) growth and the formation of islands could be due to the relief of strain etc. The result also explains why thick Te layer is needed to reach the tunneling regime as shown in Fig 2 and Fig 3.

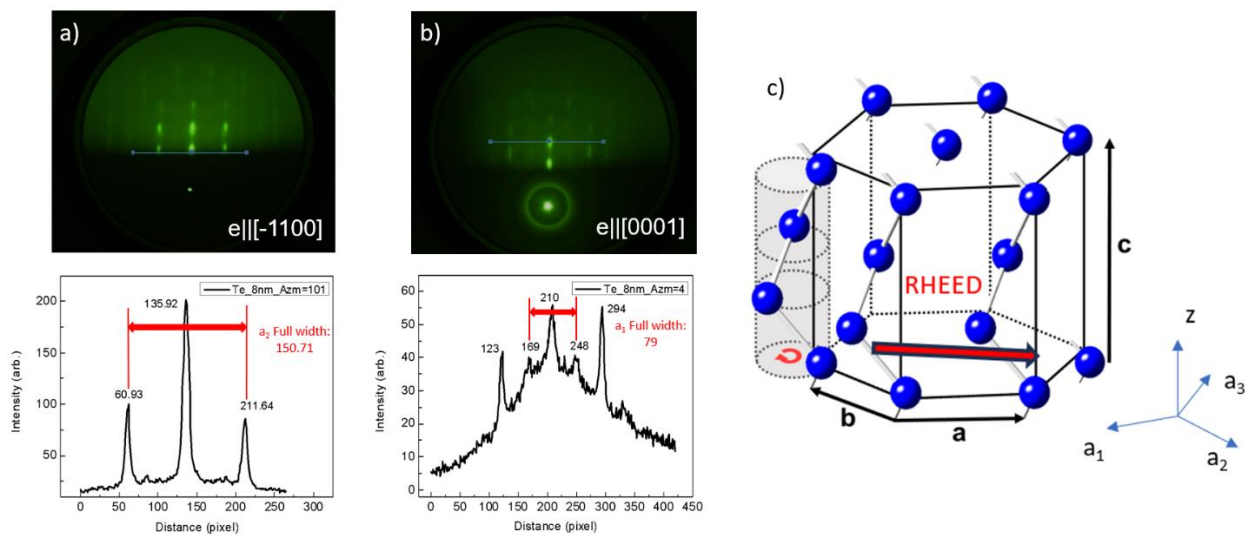

**Figure S5 | Crystaline rotation symmetry of the Te layer:** **(a)** top: the RHEED pattern taken when the electron beam is applied along the [-1100] direction of the hexagonal lattice with respect to Au(111) (see Fig S1a); bottom: the profile of a line cut along the designated line showing in the top RHEED image. The spacing of the diffraction lines is inversely proportional to the real space lattice spacing; **(b)** similar to (a) except that the electron beam is directed along the [0001] direction with respect to Au(111) (see Fig S1a); **(c)** The schematic of the trigonal Te unit cell.

Although the RHEED pattern forms dots when the Te layer is thick, the diffraction dots are arranged in diffraction lines (or streaks). Upon rotating the azimuthal angle (Fig S5), we observed a clear rotation symmetry of the diffraction lines, suggesting that the Te layer is epitaxial albeit island formation. The rotation is calibrated according to Au(111) lattice (Fig S1a), in which the periodic

RHEED patterns between $e\|[-1100]$ and $e\|[0001]$ is due to the six-fold rotation symmetry of Au(111). Here, similar rotation symmetry is observed for the Te RHEED when the electron beam is rotated from [-1100] to [0001] direction (Fig S5a, b), which suggests that Te grown on Au(111) shares similar six-fold rotation symmetry and indicates trigonal Te.

## V. Signatures of surface gap in a sample with 10 nm Au(111)/Nb (8nm)

The $dI/dV$ spectra of a sample with 10 nm Au(111) on Nb. Features reminiscent of surface superconducting gap $\Delta_S$ are observed (Fig S6b). The gap closes quickly when a planar magnetic field $\mu_0 H$ is applied (Fig S6a). When the $dI/dV$ peaks are merged at $V_{bias} \sim 0$ mV, we don't observe a ZBCP. Therefore, these peaks are more likely to be $\Delta_S$ rather than ABS. At $\mu_0 H = 0$ T, despite the thinner Au(111), $\Delta_S$ in Fig S6b is much smaller than that in Fig 2. However, we would like to point out that the size of $\Delta_S$ is not solely dependent upon the thickness of Au(111); instead, mechanisms governing surface-bulk scattering play a crucial role[18]. Therefore, we do not expect a direct relationship between $\Delta_S$ and the thickness of Au(111).

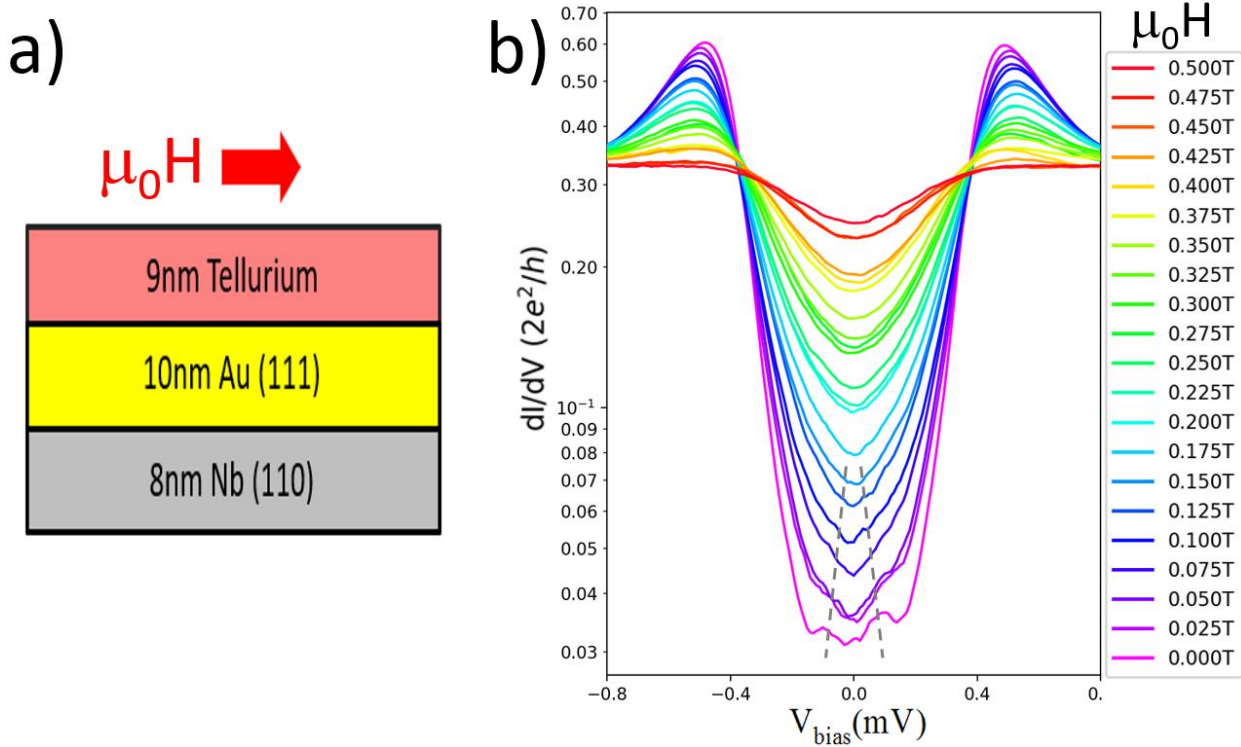

**Figure S6 | Signature of surface gap in 10 nm Au(111)/Nb (8nm): (a)** Stack layout of the layers and the direction of $\mu_0 H$; **(b)** The $dI/dV$ spectra as a function of $V_{bias}$ and $\mu_0 H$. Features reminiscent of surface superconducting gap $\Delta_S$ are observed. The dashed lines are guided by eye to show the closing of $\Delta_S$ under $\mu_0 H$.

## VI. Intrinsic quality factor $Q_i$ in standard 100 nm thick sputtered Nb films

Compared to the resonators made using 100 nm thick Nb (Fig S7), the Au(111) (5 nm)/Nb (12 nm) resonators (Fig 4e) have a comparable $Q_i$ despite the Nb layer being one order of magnitude thinner. This confirms that Au(111) protects the thin Nb layer from forming Nb oxides, which are known critical for causing microwave loss.

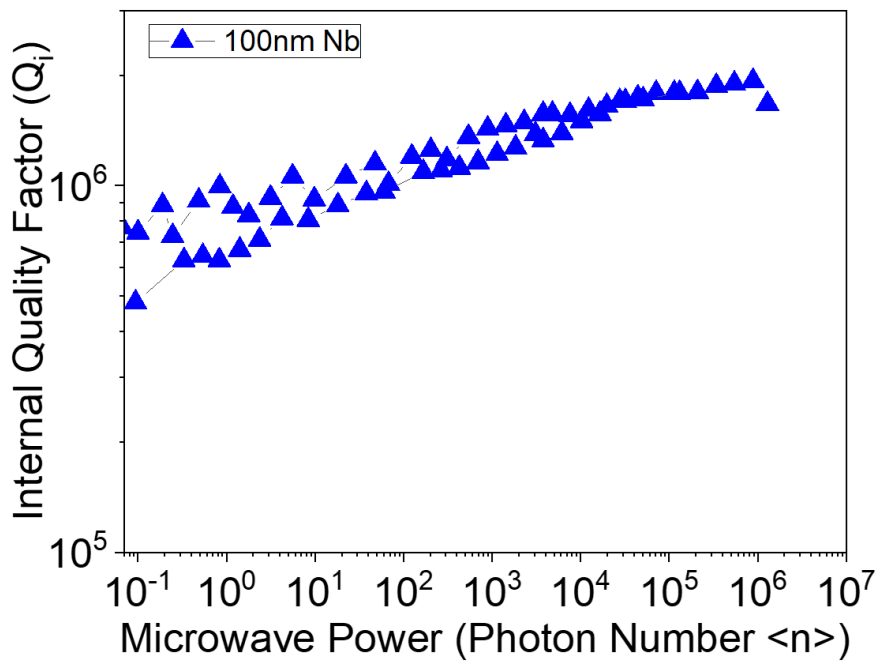

**Figure S7 | Internal quality factor $Q_i$ of a standard 100 nm Nb resonator:** $Q_i$ is comparable to the resonators made using a one order of magnitude thinner Nb layer shown in Fig 4e.